\documentclass[letter]{aa} 

\usepackage{natbib}
\usepackage{graphicx}
\usepackage{txfonts}
\begin{document}

   \title{The first Frontier Fields cluster: 4.5$\mu$m excess in a $z\sim$8 galaxy candidate in Abell 2744 }


   \author{N. Laporte \inst{1,2}
            \and
          A. Streblyanska \inst{1,2}
          \and
          B. Clement \inst{3} 
          \and
          I. P\'erez-Fournon \inst{1,2}
          \and
          D. Schaerer \inst{4,5}
          \and
          H. Atek \inst{6}
          \and
          F. Boone \inst{4,9}
          \and
          J.-P. Kneib \inst{6,7}
         \and
         E. Egami \inst{3}
         \and
          P. Mart\'inez-Navajas  \inst{1,2}
          \and
          R. Marques-Chaves  \inst{1,2}
          \and
          R. Pell\'o \inst{4,9}
          \and
          J. Richard \inst{8}
          }

   \institute{Instituto de Astrofisica de Canarias (IAC), E-38200 La Laguna, Tenerife, Spain \\
          \email{nlaporte@iac.es, alina@iac.es, ipf@iac.es, paloma@iac.es, rmarques@iac.es}
         \and
             Departamento de Astrofisica, Universidad de La Laguna (ULL), E-38205 La Laguna, Tenerife, Spain \\
             \and
             Steward Observatory, University of Arizona, 933 N. Cherry Ave, Tucson, AZ 85721, USA \\
             \email{bclement@as.arizona.edu, eegami@as.arizona.edu}
             \and
             IRAP, CNRS - 14 Avenue Edouard Belin - F-31400 Toulouse, France \\
             \email{roser.pello@irap.omp.eu, frederic.boone@irap.omp.eu}
             \and
             Observatoire de Gen\'eve - 51 Chemin des Maillettes - CH-1290 Sauverny - Switzerland \\
             \email{daniel.schaerer@unige.ch}
             \and
             Laboratoire d'Astrophysique, EPFL, Observatoire de Sauverny, CH-1290 Versoix, Switzerland \\
              \email{hakim.atek@epfl.ch,jean-paul.kneib@epfl.ch}
	    \and
              Aix Marseille Universit\'e, CNRS, LAM (Laboratoire d'Astrophysique de Marseille) UMR 7326, 13388, Marseille, France
             \and
            CRAL - UMR 5574 - Observatoire de Lyon - 9 avenue Charles Andr\'e - F-69561 Saint Genis Laval - France \\
            \email{johan.richard@univ-lyon1.fr}
            \and
            Universit\'e de Toulouse; UPS-OMP; IRAP; Toulouse, France\\
                                 }

   \date{Received 3 December 2013 / Accepted 31 January 2014}

 
  \abstract
  {}
   {We present in this letter the first analysis of a  $z\sim$8 galaxy candidate found in the \textit{Hubble} and \textit{Spitzer} imaging data of Abell 2744 as part of the Hubble Frontier Fields legacy program.}
   {We applied the most commonly used methods to select exceptionally high-redshift galaxies by combining non-detection and color criteria using seven HST bands. We used GALFIT on IRAC images to fit and subtract contamination of bright nearby sources.The physical properties were inferred from Spectral Energy Distribution-fitting using templates with and without nebular emission.}
   {This letter is focused on the brightest candidate we found (m$_{F160W}$=26.2) over the 4.9 arcmin$^2$ field of view covered by the WFC3. It is not detected in the ACS bands and at 3.6$\mu$m, while it is clearly detected at 4.5$\mu$m with rather similar depths. This break in the IRAC data might be explained by strong [OIII]+H$\beta$ lines at $z\sim$8 that contribute to the 4.5$\mu$m photometry. The best photo-$z$ is found at $z\sim$8.0$^{+0.2}_{-0.5}$, although solutions at low-redshift ($z\sim$1.9) cannot be completely excluded , but they are strongly disfavored by the SED-fitting. The amplification factor is relatively small at $\mu$=1.49$\pm$0.02. The star formation rate in this object ranges from 8 to 60 M$_{\odot}$.yr$^{-1}$, the stellar mass is in the order of M$^{\star}$=(2.5-10) $\times$10$^{9}$ M$_{\odot}$, and the size is $r\approx$0.35$\pm$0.15 kpc.  }
   {This object is one of the first $z\sim$8 Lyman Break Galaxy candidates showing a clear break between 3.6$\mu$m and 4.5$\mu$m which is consistent with the IRAC properties of the first spectroscopically confirmed galaxy at a similar redshift. Due to its brightness, the redshift of this object could potentially be confirmed by near infrared spectroscopy with current 8-10m telescopes. The nature of this candidate will be revealed in the coming months with the arrival of new ACS and Spitzer data, increasing the depth at optical and near-IR wavelengths.}


   \maketitle
%

\section{Introduction}
One of the main topics of modern astronomy is studying the early Universe through the observation of the first luminous objects. In the past ten years, several attempts have been made to constrain the properties of the Universe during the first billion years, and many surveys and instruments have been designed to push the boundaries of the observable Universe, such as CLASH \citep{2012ApJS..199...25P}, HST Frontier Fields \footnote{www.stsci.edu/hst/campaigns/frontier-fields/}, KMOS$/$ESO \citep{2006SPIE.6269E..44S}, or MOSFIRE$/$Keck \citep{2012SPIE.8446E..0JM} . These new facilities have helped to strongly increase the number of very high-redshift candidates ($z\geq$8.0) from a dozen in 2006 \citep{2006Natur.443..189B} to a hundred in 2013 (e.g. \citealt{2012ApJ...760..108B}, \citealt{2012ApJ...759..135O}). However, to date, only two objects among all the $z\geq$ 7.5 candidates have  been confirmed by spectroscopy (\citealt{2013Natur.502..524F} - F13 hereafter- and \citealt{2009Natur.461.1254T}), but none at $z\geq$8.5 . Most of the photometric candidates are too faint to be confirmed spectroscopically with current spectrographs, and require the arrival of future telescopes, such as the JWST\footnote{webpage : www.jwst.nasa.gov/}, the E-ELT \footnote{www.eso.org/public/teles-instr/e-elt/}, or the TMT\footnote{www.tmt.org }.  
 Two complementary approaches can be used to select the highest redshift objects: deep blank fields, to select the brightest objects (BoRG survey, \citealt{2011ApJ...727L..39T}; WUDS survey, Pello et al. in prep), or lensing fields, to benefit from the cluster magnification on a background source (\citealt{2011A&ARv..19...47K}, \citealt{2012Natur.489..406Z}, \citealt{2012ApJ...745..155H}).

The Frontier Fields survey proposes to combine the two previous methods by observing six lensing clusters and six blank fields to a depth similar to the \textit{Hubble Ultra Deep Field} (5$\sigma \sim$29AB from F435W to F160W). Observations will be spread over three years, starting in October 2013. The first high-level products have been released on November 1 for the cluster Abell 2744, and a first analysis of these data has been published by \citet{2013arXiv1311.7670A}. In this letter we focus on the brightest $z\sim$8 candidate selected in this lensing cluster. 
In section \ref{data} we present the dataset we used and the data reduction steps we followed to produce high-level quality IRAC data. In section \ref{LBG}, we explain the method we used to select high-redshift galaxy candidates. The $z\sim$8 galaxy candidate and its physical properties are presented and discussed in section \ref{properties}. 

The concordance cosmology is adopted throughout this paper, with $\Omega_{\Lambda}$=0.7, $\Omega_{m}$=0.3 and H$_0$=70 km s$^{-1}$ Mpc$^{-1}$. All magnitudes are quoted in the AB system \citep{1983ApJ...266..713O}.

\section{Data properties}
\label{data}

\subsection{HST data}
This work is based on the December 16  \textit{Space Telescope Science Institute}\footnote{data  : http://archive.stsci.edu/prepds/frontier/} release, which includes 100\% of the WFC3 data planned around the lensing cluster Abell 2744. ACS data come from the HST archive as part of program ID: 11689 (PI : R. Dupke) and consist of six orbits in F435W and five orbits each in F606W and F814W. WFC3 data combined two datasets: one from a supernovae search in that field (ID: 13386, PI: S. Rodney) and the Frontier Fields program (ID: 13495, PI: J. Lotz). The depth of each image was computed using noise measurements in thousands of empty 0.4'' diameter apertures that did not overlapp in a picture where bright sources were masked (cf. Table \ref{data_properties})

\subsection{Spitzer data}
Mid-infrared \textit{Spitzer} imaging of Abell 2744 using the Infrared Array Camera (IRAC) was obtained on September 2013 as part of the Frontier Field Spitzer program (PI: T. Soifer). The dataset we used represents 50\% of the planned observations for the cluster. We used corrected Basic Calibrated Data (cBCD) images that are provided by the Spitzer Science Center (SSC) and are automatically corrected by pipeline for various artifacts such as muxbleed, muxstripe, and pulldown. The cBCD frames and associated mask and uncertainty images  were processed, drizzled (with a factor of 0.65), and combined into final mosaics using the standard SSC reduction software MOPEX. The mosaics have a 3-sigma point source sensitivity (measured from the noise in a clean area) of 0.139 $\mu$Jy and  0.225 $\mu$Jy in 3.6$\mu$m and 4.5$\mu$m, respectively.


\begin{table}
\caption{Properties of \textit{HST} and \textit{Spitzer} data.}             
\label{data_properties}      
\centering                          
\begin{tabular}{c | c c c c l}        
\hline\hline                 
Filter & $\lambda_{central}$ & $\Delta\lambda$ & t$_{exp}$ & m(5$\sigma$) & Instrument \\    
         &  [$\mu$m]  & [$\mu$m]                   &   [ks]      &                         &                  \\          
 \hline                        
 F435W	& 0.431 & 0.073   & 16.16 &  27.4 & ACS \\
 F606W	& 0.589 & 0.156 & 13.25 &  28.0 & ACS \\
 F814W	& 0.811 & 0.166 & 13.25 &  27.1 & ACS \\
 F105W	& 1.050 &	0.300 & 46.52 & 28.6 & WFC3 \\
 F125W	& 1.250 &	0.300 & 16.32 & 28.5 & WFC3 \\
 F140W	& 1.400 & 0.400 & 22.43 & 28.7 & WFC3 \\
 F160W	& 1.545 & 0.290 & 46.57 & 28.2 & WFC3 \\
\hline
 3.6	& 3.550 & 0.750   &90.9$^{a}$ & 0.139$^{b}$ & IRAC \\
 4.5 	& 4.493 & 1.015 & 90.9$^{a}$ & 0.225$^{b}$ & IRAC \\
\hline
\hline                                   
\end{tabular}
\tablefoot{(1) Filter ID, (2) filter central wavelength, (3) filter FWHM, (4) exposure time, (5) depth at 5$\sigma$ in a 0.4'' diameter aperture, (6) instrument.\\ 
$^a$ on source \\
$^b$ 3$\sigma$ point source sensitivity in $\mu$Jy}
\end{table}
%
%
\section{Selection of $z\sim$8 Lyman break galaxies}
\label{LBG}
\begin{figure*}
   \centering
           \includegraphics[width=13cm]{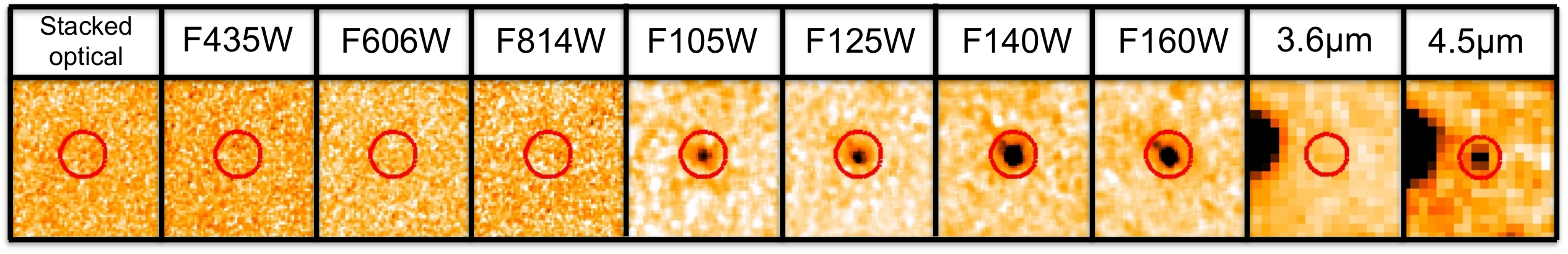}
      \caption{ Postage stamps of the Y$_{105}$-dropout. The size of each HST stamp is 2''x2'' (7''x7'' for the IRAC channels) and the position of the target is displayed by a red circle of 0.4'' (ACS and WFC3) and 1.4'' (IRAC) radius. We also show the mean stacked optical image computed using the three ACS bands.   }
         \label{stamps}
   \end{figure*}

The most successful method for selecting high-redshift galaxies is the Lyman break technique \citep{1999ApJ...519....1S}, which combines non-detection/detection criteria with color selection. We used SExtractor \citep{1996A&AS..117..393B} in double-image mode using a weighted sum of F105W and F125W images as detection. The extraction parameters were defined to maximize the number of detections. The catalog contains $\sim$ 6600 detections. The non-detection criteria we applied are 
\begin{equation}
\centering
m_{F435W},m_{F606W}, m_{F814W}>m(2\sigma) \text{    } \cup \text{    }
m_{F125W}<m(10\sigma).
\end{equation}
Therefore, we limited our search to the bright objects where a F814W-F125W break larger than 1.5mag should be observed to limit the selection of mid-$z$ interlopers \citep{2012MNRAS.425L..19H}. We computed the color window by using templates from \citet{2003MNRAS.344.1000B}, \citet{1980ApJS...43..393C}, \citealt{1996ApJ...467...38K}, \citet{2007ApJ...663...81P}, and \citet{1998ApJ...509..103S}. The color criteria we adopted were the following :
\begin{align}
m_{F105W}-m_{F125W} &>0.6 \\
m_{F125W}-m_{F160W} &<0.6 \\
m_{F105W}-m_{F125W} &>4.6\times (m_{F125W}-m_{F160W}) -0.8 .
\end{align}
Colors were measured on psf-matched data (matched to the poorest resolution) 
using Kron-like apertures. To obtain the total magnitude in each of the WFC3 bands, we applied aperture corrections using F160W MAG\_AUTO following the method described in F13. Among the 21  detections fulfilling the Lyman Break Technique criteria, only two objects were retained after visual inspection. This last step is crucial to remove spurious (e.g. all detections in the halo of bright galaxies) and very faint low-redshift objects for which the large optical error bars explain why they entered our selection criteria. We focus on the brightest candidate (RA$_{J2000}$ :  00:14:25.083 -- DEC$_{J2000}$: -30:22:49.70). As in F13, we noted that this candidate seems to have a clumpy morphology, with a faint source 0.25'' away from the main one (see Figure \ref{stamps}).

%
%

\begin{table}
\caption{Photometry of the $z\sim$8 galaxy candidate}             
\label{z8_phot}      
\centering           
\scriptsize               
\begin{tabular}{c | c c c c c c c c cl}        
\hline\hline                 
Filter & F105W & F125W 	& F140W & F160W & 3.6$\mu$m & 4.5$\mu$m\\    
         		&		&		&		&                  \\          
 \hline                        
Abell2744\_Y1 	&	27.50	&	26.32	&	26.26	&     26.25   &  >25.48$^{a}$   &  25.16     \\
		  	&   $\pm$0.08	&   $\pm$0.04	&   $\pm$0.03	&     $\pm$0.04	 &    & $\pm$0.16\\ 
	
\hline
\hline                                   
\end{tabular}
\tablefoot{Kron-like aperture corrected photometry . Error bars are computed using noise measured in empty apertures around the object.\\
$^{a}$ 3$\sigma$ limit at the position of our candidate.}
\end{table}
%
%
Owing to a bright source in the vicinity of our object on the IRAC data ($\sim$ 4.4'' from the $z\sim$8 candidate), we decided to perform a few additional steps to verify the Spitzer photometry.
We used the galaxy-fitting program GALFIT \citep{2002AJ....124..266P} with our own IRAC PSFs to fit and subtract all nearby sources around our object in a manner similar to other high-redshift IRAC studies (e.g., F13). Then, we measured the photometry in a residual (contamination-free) image in a 1.4'' radius aperture.
At 3.6$\mu$m, the source is un-detected in original and in residual images. If we consider the 4.5$\mu$m detection and the 3.6$\mu$m 3 sigma limit, the flux ratio is higher than $\approx$1.3.

%
%

\section{Physical properties}
\label{properties}

\subsection{Photometric redshift}
Photometric redshifts were computed with a new version (v12.2) of the public code {\it Hyperz\/} ({\it New$-$Hyperz\/}\footnote{http://userpages.irap.omp.eu/$\sim$rpello/newhyperz/}), originally developped by \citet{2000A&A...363..476B}. The method consists of fitting the Spectral Energy Distribution (SED) with a library of 14 templates: 8 evolutionary synthetic SEDs extracted from \citet{2003MNRAS.344.1000B}, with Chabrier IMF \citep{2003PASP..115..763C} and solar metallicity ; a set of 4 empirical SEDs compiled by \citet{1980ApJS...43..393C}, and 2 starburst galaxies from the \citet{1996ApJ...467...38K} library.  In case of non-detection in a given band, the flux in this band was set to zero, with an error bar corresponding to the limiting flux.

\begin{figure}
   \centering
	      \includegraphics[width=7cm]{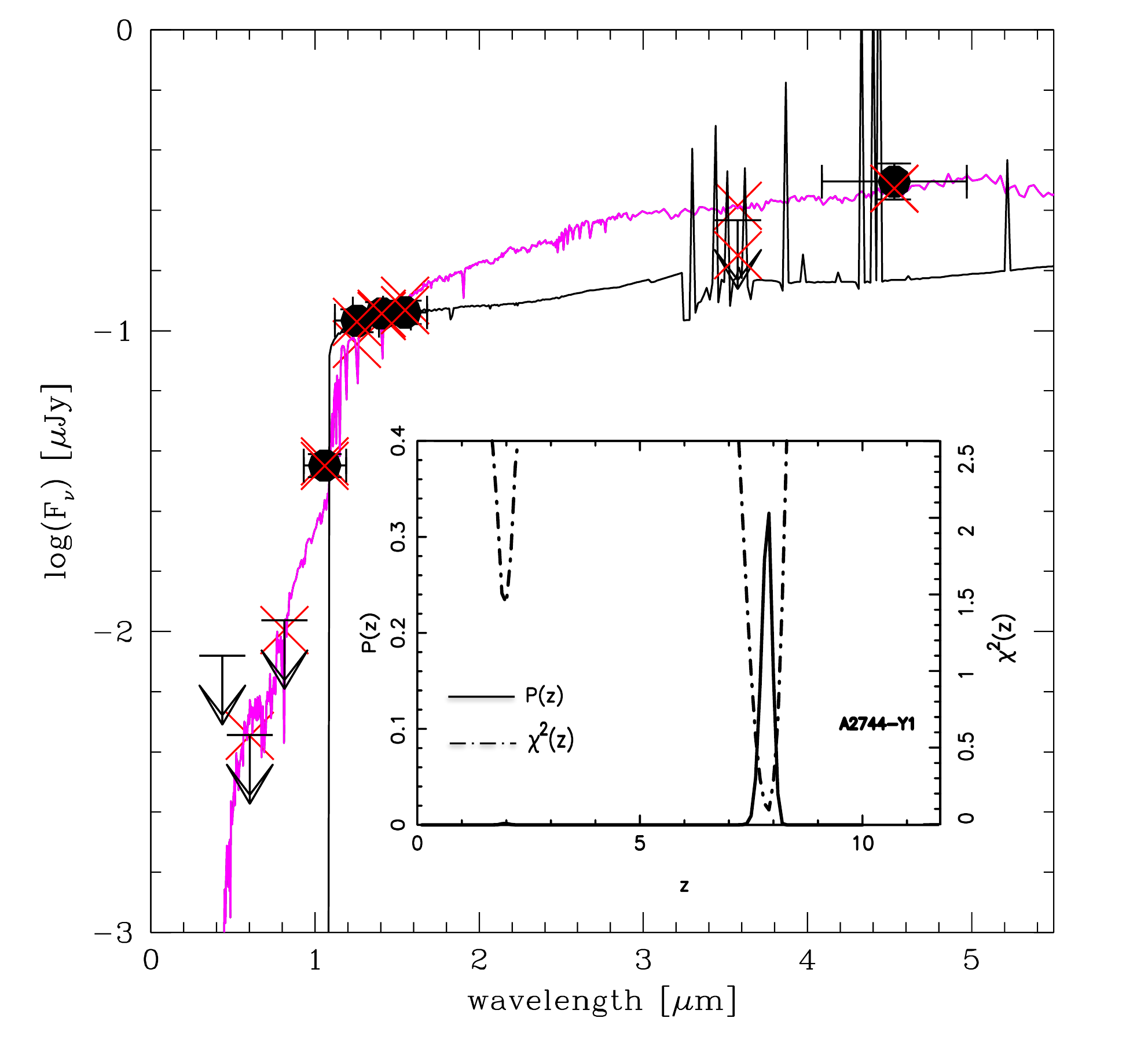}
      \caption{ Fit of the $z\sim$8 galaxy candidate SED at high- (black line) and low-redshift (magenta line). ACS upper limits are shown at 1$\sigma$ and the IRAC non-detection is plotted at 3$\sigma$. The high-$z$ SED fit shown here is for 1/50 solar metallicity, imposing an age prior of $>50$ Myr. The high-redshift solution shows an excess at 4.5$\mu$m due to [OIII] and H$\beta$ emission lines, as already observed in the $z\sim$7.51 galaxy published in F13. P($z$) and $\chi^2$($z$) are also plotted.  }
         \label{SED}
   \end{figure}

 The best SED-fit is found at $z\sim$7.98 ($\chi^{2}_{red}$=0.17), with 1$\sigma$ confidence interval ranging from  $z\sim$7.5 to 8.2. We also  fitted the SED assuming a low-redshift solution, with $z$ ranging from 0.0 to 3.0.   In that configuration the best SED-fit is found at $z\sim$1.92 ($\chi^2$=1.17 - 1$\sigma$ : 1.7 - 2.1). This $z\sim$8 candidate has an SED remarkably similar to the F13 $z\sim$7.51 galaxy where a strong break is observed in the IRAC data. 
   For these two galaxies, the excess of flux observed at 4.5$\mu$m might be explained by contamination from strong H$\beta$+[OIII] lines. For the SED fit shown in Fig.\ref{SED} obtained assuming  1/50 solar metallicity and imposing an age prior of $>$50 Myr, the  [OIII]5007, H$\beta$ restframe equivalent widths are 600 and 190 \AA\ , respectively, which are compatible with the values derived by \citet{2013arXiv1307.5847S} from the photometry of seven $z\sim$ 6.6-7 Lyman Break Galaxies (LBG). 
   At this stage and even if the high-redshift solution seems more likely and the low-$z$ solution disfavored by the SED-fitting, it appears difficult to conclude on the nature of this source. The arrival of new IRAC data in 2014 will improve the signal to noise ratio at 3.6$\mu$m and help to better understand the nature of this object.

%

\subsection{Magnification}
Part of the interest in using lensing by galaxy clusters to search for very high-redshift galaxies is the magnification by the cluster lens. However, this object is relatively far from the cluster core.  We estimated an amplification factor of $\mu$=1.49$\pm$0.02 using the public lensing model provided by the CATS group (Richard et al., in prep) in the framework of the Frontier Fields. This factor is consistent with those found using other lensing models produced by Merten ($\mu$=1.50), Sharon ($\mu$=1.91), Williams ($\mu$=1.16), and Zitrin ($\mu$=1.33-2.11), confirming a moderate amplification regime for this object.  

\subsection{Star formation rate, mass and size}

In this section, the Star Formation Rates (SFR), mass and luminosities are corrected for magnification, and are derived assuming a Salpeter IMF from 0.1--100 M$_{\odot}$. Overall, the quantities derived from SED fits are fairly uncertain, since they depend on assumptions on the metallicity and degeneracies in age--extinction. We therefore only give indicative values for these quantities.

 With an absolute UV magnitude M$_{1500}$ = -20.5 the star formation rate is SFR $\approx$ 8 M$_{\odot}.yr^{-1}$  using the standard \citet{1998ARA&A..36..189K} relation, and without correcting for attenuation by dust. Standard SED fits with solar metallicity models and neglecting nebular emission yield SFR $\sim$ 10 M$_{\odot}.yr^{-1}$ for a $A_V$=0.15. When nebular emission is included, following the models of Schaerer \& de Barros (2009, 2010), the best fits yield SFR $\sim$ 8--60 M$_{\odot}.yr^{-1}$, depending on the adopted metallicity and the minimum age imposed. 
The extinction ranges from $A_V \sim 0.05-0.8$, and the expected IR luminosity of this galaxy from $\log(L_{IR}) \sim$ 9.7--11.4, as predicted from the amount of energy absorbed in the UV.

The estimated stellar mass is on the order of $M_{\star} \sim$ (2.5-10) $\times$ 10$^9$ M$_{\odot}$ when adopting an age prior of $t>$50 Myr. If younger ages are allowed, which is favored by some best fits, the mass may be a factor 3 lower. Standard SED fits neglecting nebular emission give the largest mass, M$_{\star} \sim 1\times 10^{10}$ M$_{\odot}$. In conclusion, whereas standard models yield a specific SFR, sSFR $\sim$ 1 Gyr$^{-1}$, the sSFR is $\sim$ 20 times higher with nebular emission (for ages $t>50$ Myr), and even higher than that without age constraint. These values are comparable with those derived for example by \citet{2010A&A...515A..73S}, \citet{2012arXiv1207.3663D}, and \citet{2013arXiv1307.5847S} at $z >6$. The size of this galaxy candidate was computed using the SExtractor half-light radius corrected for PSF broadening and GALFIT modeling assuming a Sersic profile. The results are similar ($r_{SE}=$0.3$\pm$0.1 kpc and $r_{GAL}=$0.35$\pm$0.15 kpc after correction for magnification) and agree well with other studies of $z\sim$8 galaxies (\citealt{2013ApJ...777..155O}, \citealt{2010ApJ...709L..21O}). 

\section{Discussions}

The most common sources of high-redshift sample contamination come from Super-Novae, active galactic nuclei, low-mass stars, photometric scatter, transient objects, spurious sources, extremely red galaxies, etc. This $z\sim$8 candidate is detected in data collected with two instruments (\textit{HST} and \textit{Spitzer}), which limits the contamination by spurious sources. Moreover, the observations were spread over several weeks, allowing us to remove variable and transient sources from our sample. The colors of this source are inconsistent with the colors expected for L,T dwarfs (see figure 1 of \citealt{2010ApJ...709L.133B}). Moreover, the SExtractor stellarity parameter (CLASS\_STAR between 0.1 and 0.2) and the GALFIT work presented above show that our object is resolved at least in the F125W, F140W, and F160W bands, which is unlikely for stars. 

The expected number of objects in the luminosity regime of this candidate, in the field of view covered by HST and assuming the UV luminosity function published by \citet{2012ApJ...752L...5B}, is $\sim$1$\pm$1. The comoving volume explored by this survey, taking into account the amplification map of the cluster, is $\sim$3815 Mpc$^3$. Therefore the number density of $z\sim$8 galaxies is $\Phi$(M$_{1500}$=-20.47)=2.6$\times$10$^{-4}$ mag$^{-1}$Mpc$^{-3}$ (see \citealt{2013arXiv1311.7670A}, for more details) and it is consistent with recent studies published at $z\sim$8 (e.g. \citealt{2013ApJ...768..196S})

The SED of the $z\sim$8 galaxy candidate is comparable with the recent $z\sim$7.5 LBG confirmed by near-IR spectroscopy using MOSFIRE on Keck (F13) regarding the break in magnitude between optical and NIR data (F606W-F105W $>$ 2.5mag) and the detection at 4.5$\mu$m combined with the non-detection at 3.6$\mu$m with similar depth. Its colors fulfill the color criteria defined by \citet{2012ApJ...759..135O} and \citet{2011MNRAS.414.1455L} and its \textit{intrinsic} SED is  similar to the $z\sim\ $8 candidates selected in the CANDELS survey \citep{2011ApJS..197...35G}. Furthermore, the H$_{160}$-[4.5] $=$1.1 and J$_{125}$ - H$_{160}$ colors agree well with the trend observed by \citet{2013ApJ...777L..19L} for $z\sim$8 objects detected in Spitzer data. As shown in recent papers (e.g. \citealt{2013arXiv1307.5847S}, Clement et al. in prep.), a strong contribution of H$\beta$ and [OIII] emission lines at $z>$7 to the \textit{Spitzer} photometry might be another argument in favor of the high-redshift solution.

\section{Conclusions}
Within the 4.9 arcmin$^2$ field of view covered by \textit{HST} in A2744, we found a relatively bright $z\sim$8 galaxy candidate well detected at 4.5$\mu$m, but undetected at 3.6$\mu$m. This excess might be the signature of strong H$\beta$ and [OIII] emission lines in the IRAC channel 2. Based on an SED-fitting including nebular emission, this object appears to be a good candidate at $z\sim$8, and its photometric properties, such as mass, SFR and size, are consistent with expectations at $z>7$.  In the coming months, a new set of ACS and IRAC data will become available and may confirm the break observed. These data will improve the photometry and the overall SED and physical parameters, and hopefully will confirm or discard the very high-redshift nature of this source. Most of the $z\sim8$ sources known to date are too faint to be spectroscopically confirmed with NIR spectrographs, and they require the arrival of new facilities to be confirmed (e.g. JWST or the E-ELT). However, this object is bright enough to be explored with current spectrographs in a reasonable amount of time ($\approx$10-15h). If the high-redshift solution is confirmed, it will be the second source at $z>$7.5 with a confirmed break in IRAC data.

\begin{acknowledgements}
The authors thank the anonymous referee for useful comments, the HST and Spitzer directors and FF teams for their work to make these fantastic data available.This work was supported by the Spanish MINECO under projects AYA2010-21697-C05-04 and FIS2012-39162-C06-02. JPK et HA acknowledges support from the ERC advanced grant LIDA, JR  support from the ERC starting grant CALENDS. This work also received support from the French Agence Nationale de la Recherche bearing the reference ANR-09-BLAN-0234.
\end{acknowledgements}


\bibliographystyle{aa}  
\bibliography{arxiv_v2.bib} 

\begin{thebibliography}{37}
\expandafter\ifx\csname natexlab\endcsname\relax\def\natexlab#1{#1}\fi

\bibitem[{{Atek} {et~al.}(2013){Atek}, {Richard}, {Kneib}, {Clement}, {Egami},
  {Ebeling}, {Jauzac}, {Jullo}, {Laporte}, {Limousin}, \&
  {Natarajan}}]{2013arXiv1311.7670A}
{Atek}, H., {Richard}, J., {Kneib}, J.-P., {et~al.} 2013, ArXiv e-prints

\bibitem[{{Bertin} \& {Arnouts}(1996)}]{1996A&AS..117..393B}
{Bertin}, E. \& {Arnouts}, S. 1996, \aaps, 117, 393

\bibitem[{{Bolzonella} {et~al.}(2000){Bolzonella}, {Miralles}, \&
  {Pell{\'o}}}]{2000A&A...363..476B}
{Bolzonella}, M., {Miralles}, J.-M., \& {Pell{\'o}}, R. 2000, \aap, 363, 476

\bibitem[{{Bouwens} \& {Illingworth}(2006)}]{2006Natur.443..189B}
{Bouwens}, R.~J. \& {Illingworth}, G.~D. 2006, \nat, 443, 189

\bibitem[{{Bouwens} {et~al.}(2010){Bouwens}, {Illingworth}, {Oesch},
  {Stiavelli}, {van Dokkum}, {Trenti}, {Magee}, {Labb{\'e}}, {Franx},
  {Carollo}, \& {Gonzalez}}]{2010ApJ...709L.133B}
{Bouwens}, R.~J., {Illingworth}, G.~D., {Oesch}, P.~A., {et~al.} 2010, \apjl,
  709, L133

\bibitem[{{Bouwens} {et~al.}(2012){Bouwens}, {Illingworth}, {Oesch}, {Trenti},
  {Labb{\'e}}, {Franx}, {Stiavelli}, {Carollo}, {van Dokkum}, \&
  {Magee}}]{2012ApJ...752L...5B}
{Bouwens}, R.~J., {Illingworth}, G.~D., {Oesch}, P.~A., {et~al.} 2012, \apjl,
  752, L5

\bibitem[{{Bradley} {et~al.}(2012){Bradley}, {Trenti}, {Oesch}, {Stiavelli},
  {Treu}, {Bouwens}, {Shull}, {Holwerda}, \& {Pirzkal}}]{2012ApJ...760..108B}
{Bradley}, L.~D., {Trenti}, M., {Oesch}, P.~A., {et~al.} 2012, \apj, 760, 108

\bibitem[{{Bruzual} \& {Charlot}(2003)}]{2003MNRAS.344.1000B}
{Bruzual}, G. \& {Charlot}, S. 2003, \mnras, 344, 1000

\bibitem[{{Chabrier}(2003)}]{2003PASP..115..763C}
{Chabrier}, G. 2003, \pasp, 115, 763

\bibitem[{{Coleman} {et~al.}(1980){Coleman}, {Wu}, \&
  {Weedman}}]{1980ApJS...43..393C}
{Coleman}, G.~D., {Wu}, C.-C., \& {Weedman}, D.~W. 1980, \apjs, 43, 393

\bibitem[{{de Barros} {et~al.}(2012){de Barros}, {Schaerer}, \&
  {Stark}}]{2012arXiv1207.3663D}
{de Barros}, S., {Schaerer}, D., \& {Stark}, D.~P. 2012, ArXiv e-prints

\bibitem[{{Finkelstein} {et~al.}(2013){Finkelstein}, {Papovich}, {Dickinson},
  {Song}, {Tilvi}, {Koekemoer}, {Finkelstein}, {Mobasher}, {Ferguson},
  {Giavalisco}, {Reddy}, {Ashby}, {Dekel}, {Fazio}, {Fontana}, {Grogin},
  {Huang}, {Kocevski}, {Rafelski}, {Weiner}, \&
  {Willner}}]{2013Natur.502..524F}
{Finkelstein}, S.~L., {Papovich}, C., {Dickinson}, M., {et~al.} 2013, \nat,
  502, 524

\bibitem[{{Grogin} {et~al.}(2011){Grogin}, {Kocevski}, {Faber}, {Ferguson},
  {Koekemoer}, {Riess}, {Acquaviva}, {Alexander}, {Almaini}, {Ashby}, {Barden},
  {Bell}, {Bournaud}, {Brown}, {Caputi}, {Casertano}, {Cassata}, {Castellano},
  {Challis}, {Chary}, {Cheung}, {Cirasuolo}, {Conselice}, {Roshan Cooray},
  {Croton}, {Daddi}, {Dahlen}, {Dav{\'e}}, {de Mello}, {Dekel}, {Dickinson},
  {Dolch}, {Donley}, {Dunlop}, {Dutton}, {Elbaz}, {Fazio}, {Filippenko},
  {Finkelstein}, {Fontana}, {Gardner}, {Garnavich}, {Gawiser}, {Giavalisco},
  {Grazian}, {Guo}, {Hathi}, {H{\"a}ussler}, {Hopkins}, {Huang}, {Huang},
  {Jha}, {Kartaltepe}, {Kirshner}, {Koo}, {Lai}, {Lee}, {Li}, {Lotz}, {Lucas},
  {Madau}, {McCarthy}, {McGrath}, {McIntosh}, {McLure}, {Mobasher},
  {Moustakas}, {Mozena}, {Nandra}, {Newman}, {Niemi}, {Noeske}, {Papovich},
  {Pentericci}, {Pope}, {Primack}, {Rajan}, {Ravindranath}, {Reddy}, {Renzini},
  {Rix}, {Robaina}, {Rodney}, {Rosario}, {Rosati}, {Salimbeni}, {Scarlata},
  {Siana}, {Simard}, {Smidt}, {Somerville}, {Spinrad}, {Straughn}, {Strolger},
  {Telford}, {Teplitz}, {Trump}, {van der Wel}, {Villforth}, {Wechsler},
  {Weiner}, {Wiklind}, {Wild}, {Wilson}, {Wuyts}, {Yan}, \&
  {Yun}}]{2011ApJS..197...35G}
{Grogin}, N.~A., {Kocevski}, D.~D., {Faber}, S.~M., {et~al.} 2011, \apjs, 197,
  35

\bibitem[{{Hall} {et~al.}(2012){Hall}, {Brada{\v c}}, {Gonzalez}, {Treu},
  {Clowe}, {Jones}, {Stiavelli}, {Zaritsky}, {Cuby}, \&
  {Cl{\'e}ment}}]{2012ApJ...745..155H}
{Hall}, N., {Brada{\v c}}, M., {Gonzalez}, A.~H., {et~al.} 2012, \apj, 745, 155

\bibitem[{{Hayes} {et~al.}(2012){Hayes}, {Laporte}, {Pell{\'o}}, {Schaerer}, \&
  {Le Borgne}}]{2012MNRAS.425L..19H}
{Hayes}, M., {Laporte}, N., {Pell{\'o}}, R., {Schaerer}, D., \& {Le Borgne},
  J.-F. 2012, \mnras, 425, L19

\bibitem[{{Kennicutt}(1998)}]{1998ARA&A..36..189K}
{Kennicutt}, Jr., R.~C. 1998, \araa, 36, 189

\bibitem[{{Kinney} {et~al.}(1996){Kinney}, {Calzetti}, {Bohlin}, {McQuade},
  {Storchi-Bergmann}, \& {Schmitt}}]{1996ApJ...467...38K}
{Kinney}, A.~L., {Calzetti}, D., {Bohlin}, R.~C., {et~al.} 1996, \apj, 467, 38

\bibitem[{{Kneib} \& {Natarajan}(2011)}]{2011A&ARv..19...47K}
{Kneib}, J.-P. \& {Natarajan}, P. 2011, \aapr, 19, 47

\bibitem[{{Labb{\'e}} {et~al.}(2013){Labb{\'e}}, {Oesch}, {Bouwens},
  {Illingworth}, {Magee}, {Gonz{\'a}lez}, {Carollo}, {Franx}, {Trenti}, {van
  Dokkum}, \& {Stiavelli}}]{2013ApJ...777L..19L}
{Labb{\'e}}, I., {Oesch}, P.~A., {Bouwens}, R.~J., {et~al.} 2013, \apjl, 777,
  L19

\bibitem[{{Lorenzoni} {et~al.}(2011){Lorenzoni}, {Bunker}, {Wilkins},
  {Stanway}, {Jarvis}, \& {Caruana}}]{2011MNRAS.414.1455L}
{Lorenzoni}, S., {Bunker}, A.~J., {Wilkins}, S.~M., {et~al.} 2011, \mnras, 414,
  1455

\bibitem[{{McLean} {et~al.}(2012){McLean}, {Steidel}, {Epps}, {Konidaris},
  {Matthews}, {Adkins}, {Aliado}, {Brims}, {Canfield}, {Cromer}, {Fucik},
  {Kulas}, {Mace}, {Magnone}, {Rodriguez}, {Rudie}, {Trainor}, {Wang}, {Weber},
  \& {Weiss}}]{2012SPIE.8446E..0JM}
{McLean}, I.~S., {Steidel}, C.~C., {Epps}, H.~W., {et~al.} 2012, in Society of
  Photo-Optical Instrumentation Engineers (SPIE) Conference Series, Vol. 8446,
  Society of Photo-Optical Instrumentation Engineers (SPIE) Conference Series

\bibitem[{{Oesch} {et~al.}(2010){Oesch}, {Bouwens}, {Carollo}, {Illingworth},
  {Trenti}, {Stiavelli}, {Magee}, {Labb{\'e}}, \&
  {Franx}}]{2010ApJ...709L..21O}
{Oesch}, P.~A., {Bouwens}, R.~J., {Carollo}, C.~M., {et~al.} 2010, \apjl, 709,
  L21

\bibitem[{{Oesch} {et~al.}(2012){Oesch}, {Bouwens}, {Illingworth}, {Gonzalez},
  {Trenti}, {van Dokkum}, {Franx}, {Labb{\'e}}, {Carollo}, \&
  {Magee}}]{2012ApJ...759..135O}
{Oesch}, P.~A., {Bouwens}, R.~J., {Illingworth}, G.~D., {et~al.} 2012, \apj,
  759, 135

\bibitem[{{Oke} \& {Gunn}(1983)}]{1983ApJ...266..713O}
{Oke}, J.~B. \& {Gunn}, J.~E. 1983, \apj, 266, 713

\bibitem[{{Ono} {et~al.}(2013){Ono}, {Ouchi}, {Curtis-Lake}, {Schenker},
  {Ellis}, {McLure}, {Dunlop}, {Robertson}, {Koekemoer}, {Bowler}, {Rogers},
  {Schneider}, {Charlot}, {Stark}, {Shimasaku}, {Furlanetto}, \&
  {Cirasuolo}}]{2013ApJ...777..155O}
{Ono}, Y., {Ouchi}, M., {Curtis-Lake}, E., {et~al.} 2013, \apj, 777, 155

\bibitem[{{Peng} {et~al.}(2002){Peng}, {Ho}, {Impey}, \&
  {Rix}}]{2002AJ....124..266P}
{Peng}, C.~Y., {Ho}, L.~C., {Impey}, C.~D., \& {Rix}, H.-W. 2002, \aj, 124, 266

\bibitem[{{Polletta} {et~al.}(2007){Polletta}, {Tajer}, {Maraschi},
  {Trinchieri}, {Lonsdale}, {Chiappetti}, {Andreon}, {Pierre}, {Le F{\`e}vre},
  {Zamorani}, {Maccagni}, {Garcet}, {Surdej}, {Franceschini}, {Alloin},
  {Shupe}, {Surace}, {Fang}, {Rowan-Robinson}, {Smith}, \&
  {Tresse}}]{2007ApJ...663...81P}
{Polletta}, M., {Tajer}, M., {Maraschi}, L., {et~al.} 2007, \apj, 663, 81

\bibitem[{{Postman} {et~al.}(2012){Postman}, {Coe}, {Ben{\'{\i}}tez},
  {Bradley}, {Broadhurst}, {Donahue}, {Ford}, {Graur}, {Graves}, {Jouvel},
  {Koekemoer}, {Lemze}, {Medezinski}, {Molino}, {Moustakas}, {Ogaz}, {Riess},
  {Rodney}, {Rosati}, {Umetsu}, {Zheng}, {Zitrin}, {Bartelmann}, {Bouwens},
  {Czakon}, {Golwala}, {Host}, {Infante}, {Jha}, {Jimenez-Teja}, {Kelson},
  {Lahav}, {Lazkoz}, {Maoz}, {McCully}, {Melchior}, {Meneghetti}, {Merten},
  {Moustakas}, {Nonino}, {Patel}, {Reg{\"o}s}, {Sayers}, {Seitz}, \& {Van der
  Wel}}]{2012ApJS..199...25P}
{Postman}, M., {Coe}, D., {Ben{\'{\i}}tez}, N., {et~al.} 2012, \apjs, 199, 25

\bibitem[{{Schaerer} \& {de Barros}(2010)}]{2010A&A...515A..73S}
{Schaerer}, D. \& {de Barros}, S. 2010, \aap, 515, A73

\bibitem[{{Schenker} {et~al.}(2013){Schenker}, {Robertson}, {Ellis}, {Ono},
  {McLure}, {Dunlop}, {Koekemoer}, {Bowler}, {Ouchi}, {Curtis-Lake}, {Rogers},
  {Schneider}, {Charlot}, {Stark}, {Furlanetto}, \&
  {Cirasuolo}}]{2013ApJ...768..196S}
{Schenker}, M.~A., {Robertson}, B.~E., {Ellis}, R.~S., {et~al.} 2013, \apj,
  768, 196

\bibitem[{{Sharples} {et~al.}(2006){Sharples}, {Bender}, {Bennett}, {Burch},
  {Carter}, {Casali}, {Clark}, {Content}, {Davies}, {Davies}, {Dubbeldam},
  {Finger}, {Genzel}, {Haefner}, {Hess}, {Kissler-Patig}, {Laidlaw}, {Lehnert},
  {Lewis}, {Moorwood}, {Muschielok}, {F{\"o}rster Schreiber}, {Pirard}, {Ramsay
  Howat}, {Rees}, {Richter}, {Robertson}, {Robson}, {Saglia}, {Tecza},
  {Thatte}, {Todd}, \& {Wegner}}]{2006SPIE.6269E..44S}
{Sharples}, R., {Bender}, R., {Bennett}, R., {et~al.} 2006, in Society of
  Photo-Optical Instrumentation Engineers (SPIE) Conference Series, Vol. 6269,
  Society of Photo-Optical Instrumentation Engineers (SPIE) Conference Series

\bibitem[{{Silva} {et~al.}(1998){Silva}, {Granato}, {Bressan}, \&
  {Danese}}]{1998ApJ...509..103S}
{Silva}, L., {Granato}, G.~L., {Bressan}, A., \& {Danese}, L. 1998, \apj, 509,
  103

\bibitem[{{Smit} {et~al.}(2013){Smit}, {Bouwens}, {Labbe}, {Zheng}, {Bradley},
  {Donahue}, {Lemze}, {Moustakas}, {Umetsu}, {Zitrin}, {Coe}, {Postman},
  {Gonzalez}, {Bartelmann}, {Benitez}, {Broadhurst}, {Ford}, {Grillo},
  {Infante}, {Jimenez-Teja}, {Jouvel}, {Kelson}, {Lahav}, {Maoz}, {Medezinski},
  {Melchior}, {Meneghetti}, {Merten}, {Molino}, {Moustakas}, {Nonino},
  {Rosati}, \& {Seitz}}]{2013arXiv1307.5847S}
{Smit}, R., {Bouwens}, R.~J., {Labbe}, I., {et~al.} 2013, ArXiv e-prints

\bibitem[{{Steidel} {et~al.}(1999){Steidel}, {Adelberger}, {Giavalisco},
  {Dickinson}, \& {Pettini}}]{1999ApJ...519....1S}
{Steidel}, C.~C., {Adelberger}, K.~L., {Giavalisco}, M., {Dickinson}, M., \&
  {Pettini}, M. 1999, \apj, 519, 1

\bibitem[{{Tanvir} {et~al.}(2009){Tanvir}, {Fox}, {Levan}, {Berger},
  {Wiersema}, {Fynbo}, {Cucchiara}, {Kr{\"u}hler}, {Gehrels}, {Bloom},
  {Greiner}, {Evans}, {Rol}, {Olivares}, {Hjorth}, {Jakobsson}, {Farihi},
  {Willingale}, {Starling}, {Cenko}, {Perley}, {Maund}, {Duke}, {Wijers},
  {Adamson}, {Allan}, {Bremer}, {Burrows}, {Castro-Tirado}, {Cavanagh}, {de
  Ugarte Postigo}, {Dopita}, {Fatkhullin}, {Fruchter}, {Foley}, {Gorosabel},
  {Kennea}, {Kerr}, {Klose}, {Krimm}, {Komarova}, {Kulkarni}, {Moskvitin},
  {Mundell}, {Naylor}, {Page}, {Penprase}, {Perri}, {Podsiadlowski}, {Roth},
  {Rutledge}, {Sakamoto}, {Schady}, {Schmidt}, {Soderberg}, {Sollerman},
  {Stephens}, {Stratta}, {Ukwatta}, {Watson}, {Westra}, {Wold}, \&
  {Wolf}}]{2009Natur.461.1254T}
{Tanvir}, N.~R., {Fox}, D.~B., {Levan}, A.~J., {et~al.} 2009, \nat, 461, 1254

\bibitem[{{Trenti} {et~al.}(2011){Trenti}, {Bradley}, {Stiavelli}, {Oesch},
  {Treu}, {Bouwens}, {Shull}, {MacKenty}, {Carollo}, \&
  {Illingworth}}]{2011ApJ...727L..39T}
{Trenti}, M., {Bradley}, L.~D., {Stiavelli}, M., {et~al.} 2011, \apjl, 727, L39

\bibitem[{{Zheng} {et~al.}(2012){Zheng}, {Postman}, {Zitrin}, {Moustakas},
  {Shu}, {Jouvel}, {H{\o}st}, {Molino}, {Bradley}, {Coe}, {Moustakas},
  {Carrasco}, {Ford}, {Ben{\'{\i}}tez}, {Lauer}, {Seitz}, {Bouwens},
  {Koekemoer}, {Medezinski}, {Bartelmann}, {Broadhurst}, {Donahue}, {Grillo},
  {Infante}, {Jha}, {Kelson}, {Lahav}, {Lemze}, {Melchior}, {Meneghetti},
  {Merten}, {Nonino}, {Ogaz}, {Rosati}, {Umetsu}, \& {van der
  Wel}}]{2012Natur.489..406Z}
{Zheng}, W., {Postman}, M., {Zitrin}, A., {et~al.} 2012, \nat, 489, 406

\end{thebibliography}

\end{document}